# The mechanism of doping and the features of phase diagrams of HTSC cuprates and ferropnictides

Kirill Mitsen[*] and Olga Ivanenko

Lebedev Physical Institute, Leninsky pr., 53 Moscow, 119991, Russia

We propose a generalized model of electronic structure modification in HTSC cuprates and ferropnictides under doping. In this model the role of doping consists in only a local change in the electronic structures of the parent phases of cuprates and ferropnictides due to the formation of trion complexes comprising a doped carrier localized in unit cell and charge transfer (CT) excitons around it. These CT excitons emerge in $CuO_4$ or $AsFe_4$ plaquettes in the $CuO_2$ or FeAs basal planes (CT plaquettes) under the influence of doped carrier, restricting its itinerancy. As the dopant concentration is increased, CT plaquettes combine into clusters of the so called CT phase. It is this CT phase that is related in the model to the HTSC phase. In support of this assumption, we determined the ranges of dopant concentrations conforming to the existence of percolation clusters of the CT phase; these ranges were shown to coincide with the positions of the superconducting domes on the phase diagrams of these compounds. The model also perfectly describes subtle features of the phase diagrams of various cuprates and ferropnictides including the "1/8" anomaly, narrow peaks in the dependences of the London penetration depth on the concentration of the dopant, and other specific features. The mechanism of the generation of free carriers in the CT phase, provided by intrinsic self-doping, was considered. The mechanism is not directly related to external doping, but is due to the interaction of band electrons with so called Heitler–London (HL) centres inherently existing in the percolation cluster of CT phase and representing pairs of adjacent $CuO_4$ or $AsFe_4$ CT plaquettes in the $CuO_2$ or FeAs basal planes. Material in CT phase was shown to represent a medium, in which the mechanism of excitonic superconductivity, specified by the interaction of band electrons with HL centres, can be realized.

---

[*] Correspondence to [mitsen@lebedev.ru]

**Introduction**

To date, there are two large families of high-temperature superconductors, cuprates and ferropnictides, which possess the most high superconducting-transition temperatures under ordinary conditions. Compounds of both classes are layered, and it is generally accepted that the quasi-two dimensional atomic planes of $CuO_2$ and FeAs are responsible for superconductivity in them. As for the nature of the normal state and the mechanism of high-$T_c$ superconductivity in these materials, these issues are yet far from being solved and are actively debated. We believe that one of the reasons for this situation is the use of a simplified doping scheme (as in the semiconductor) when constructing theoretical models, which, in our mind, does not quite adequately describe the process of the generation of free carriers in these materials.

Here we would like to propose a mechanism for the heterovalent and isovalent doping common of both classes of materials, in which the generation of free carriers proceeds as if in two stages. In the case of heterovalent doping, at the first stage, introduction of dopant atoms leads only to a local modification of the electronic structure of cuprates and pnictides, and to the formation of separate trion complexes. Such a complex consists of a doped carrier, which self-localizes in the nearest vicinity of the dopant due to charge-transfer (CT) excitons emerging under its impact. These excitons are generated in $CuO_4$ and $FeAs_4$ plaquettes bordering with the doped carrier localization region (CT plaquettes). At the second stage, with the dopant concentration rising, these CT plaquettes are joined into clusters of the so called CT phase, which is in fact a CT-exciton insulator, where free carriers are generated owing to the self-doping mechanism specific for this phase. That is, doped carriers proper (from dopants) are localized, and their role is only to form clusters of the CT phase, in which pairs of adjacent CT plaquettes (the so called HL centres) play the role of donors or acceptors. In the case of isovalent doping, embedding of dopant atoms is also accompanied with the deformation of the local electronic structure leading within some concentration range to the formation of CT phase clusters.

It is this CT phase that is related in the model to the HTSC phase. Herewith, the range of concentrations where a CT-phase percolation cluster can exist corresponds to the superconducting dome on the phase diagrams of cuprates and pnictides.

The proposed model makes it possible to explain on a uniform basis many properties of cuprates and pnictides, including the wonderful features of their superconducting phase diagrams, spatial inhomogeneity of their electronic properties, pseudogap anomalies in cuprates and other features [1]. Besides, the model considered admits the possibility of implementing in cuprates and pnictides of a specific mechanism of excitonic superconductivity potentially capable of providing for high $T_c$ [2].

## 1 Modification of band structures of cuprates and ferropnictides under doping

As is known, the predominant majority of undoped cuprates and pnictides are not superconductors. Superconductivity in them emerges as the result of heterovalent doping, i.e., at a partial substitution of an atom for another atom with higher or lower valency.

Figure 1a,b shows an arrangement of anions and cations (their projections, to be more exact) in the basal planes of cuprates and ferropnictides. In cuprates, Cu and O ions are in the same plane; in ferropnictides, Fe ions are in the plane and As ions are at the vertices of the regular tetrahedra such that their projections form a square sublattice in the basal plane.

In an undoped state, cuprates and ferropnictides demonstrate essentially different properties. Undoped cuprates are antiferromagnetic Mott insulators in which the empty subband of copper $3d^{10}$ states is separated from the occupied oxygen O2p band by a gap $\Delta_{ib} \sim 2$ eV (Fig. 1c). In undoped ferropnictides that are antiferromagnetic (bad) metals, states on the Fermi surface are formed mainly by Fe orbitals, whereas electron states on As are ~2 eV lower than $E_F$ [3] (Fig. 1d). However, despite the seeming difference in the electronic structures of undoped cuprates and ferropnictides, there is something that unites them. This can be seen if we present the electronic structure of ferropnictides in hole representation (Fig. 1e) and consider the transition of an electron or a hole between the occupied upper and empty lower state.

It is seen (Fig. 1c,e) that, to transfer an electron from an $O^{2-}$ anion to a $Cu^{2+}$ cation in cuprates (to form a $Cu^+$ ion), approximately the same energy of ~2 eV should be spent as to transfer a hole from a Fe cation (from $E_F$) to an $As^{3-}$ anion in ferropnictides (to form an $As^{2-}$ ion). An important fact, which exphazises the similarity of these compounds (and will be important for further consideration), is that the orbitals of the nearest $Cu^{1+}$ ions (in cuprates) do not overlap, as neither do the orbitals of $As^{2-}$ ions (in ferropnictides), and no chemical (covalent or metal) bond forms between them [4]. At the same time the orbitals of the nearest ions of O or Fe do overlap.

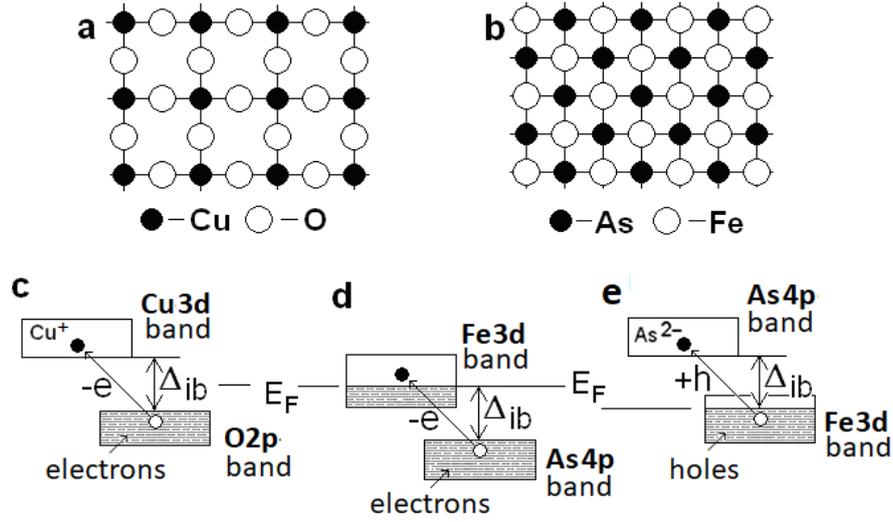

**Figure 1 a,b**, Arrangement of the projections of anions and cations in the basal planes of cuprates (**a**) and ferropnictides (**b**); **c,d**, the band structures of undoped cuprates and ferropnictides in electron representation; **e**, the band structure of undoped ferropnictides in hole representation. Energy $\Delta_{ib}$ required for interband transition related to the transfer of an electron from an oxygen ion to a copper ion (in cuprates), or to the transfer of a hole from an iron ion to an arsenium ion (in ferropnictides), is approximately the same in both cases and makes ~2 eV.

Thus, the energy required for electron transfer from an oxygen ion to a copper ion (in cuprates) or the transfer of a hole from an iron ion to an arsenium ion (in ferropnictides) is in both cases approximately the same, $\Delta_{ib} \sim 2$ eV. At the same time, exciton-like excitation is also possible, which has a lower energy $\Delta_{ct} < \Delta_{ib}$ and corresponds to the local transfer of an electron (hole) from an anion (cation) to the nearest cation (anion) (Fig. 2a,b) to form the bound state (of a CT exciton).

The events corresponding to the generation of a CT exciton are, in cuprates, the occurrence of an electron on the central Cu cation and a hole distributed over four surrounding anions of O; in ferropnictides, a hole on the As anion and an electron distributed over four surrounding cations of Fe. This hydrogen-like ionic complex, for which the condition $\Delta_{ct} = 0$ is satisfied and in which a CT exciton, resonantly interacting with the band states, can be formed, will be called a CT plaquette. In this complex, the state of a 2*p*-electron (in cuprates) and a 3*d*-hole (in ferropnictides) should be considered as a superposition of band and exciton states. In a CT phase of cuprates (ferropnictides) each Cu (As) ion is the centre of a CT plaquette.

CT excitons can form in cuprates and pnictides owing to the following features:

(1) *Low concentration of charge carriers*. Even at an optimal doping, the carrier concentration in cuprates and ferropnictides is lower than $10^{22}$ cm$^{-3}$, which corresponds to an average distance of $r_s > 0.4$ nm between the carriers and exceeds the distance between the anion and cation. This means that the interaction inside the cell is essentially unshielded, which enables the existence of well-defined CT excitons [5].

(2) *High ionicity of cuprates and ferropnictides*, which suggests a large contribution of Madelung volume energy $E_M$ to the electronic structure of the basal planes and the possibility to locally change the electronic structure by doping with localized carriers.

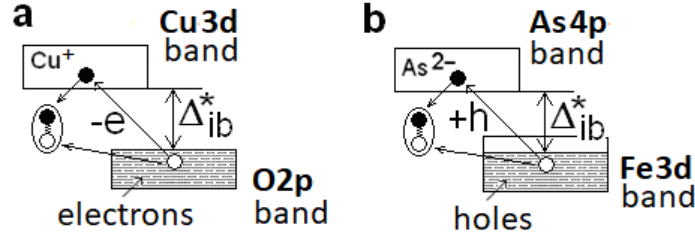

**Figure 2** Formation of a CT exciton in cuprates (**a**) and pnictides (**b**). To form a CT exciton, the band gap $\Delta_{ib}$ should be reduced to $\Delta_{ib} = \Delta_{ib}^* < E_{ex}$, where $E_{ex}$ is CT exciton binding energy.

Therefore, if $\Delta_{ib}$ is somehow gradually decreased, we shall first arrive to a state with $\Delta_{ib} = \Delta_{ib}^*$, (Fig. 3c,d) in which the charge-transfer gap $\Delta_{ct} = 0$ (i.e., $D_{ib}^* \leq E_{ex}$, where $E_{ex}$ is CT exciton binding energy). If $\Delta_{ib}$ is decreased homogeneously, the condition $\Delta_{ct} = 0$ will be satisfied for the entire basal plane. If we locally suppress $\Delta_{ib}$ in some regions, a continuous cluster of the phase with $\Delta_{ct} = 0$ will emerge at an excess of the percolation threshold over those regions.

Let us have a phase with $\Delta_{ct} = 0$ (i.e., $\Delta_{ib}^* \leq E_{ex}$). This phase (Fig. 3c,d) we will call the CT phase. Because $\Delta_{ib}^* < E_{ex}$, electrons in cuprates occupying the states in the upper part of the O2p band (Fig. 3c) will pass into $Cu^+/O^-$ exciton states, where $O^-$ denotes a hole on the oxygen neighbouring with the copper ion (Fig. 3e). Correspondingly, the emerging oxygen holes will occupy $O^-/Cu^+$ states higher than the Fermi level [6]. This transition should be considered as a transition into the state of a CT exciton insulator.

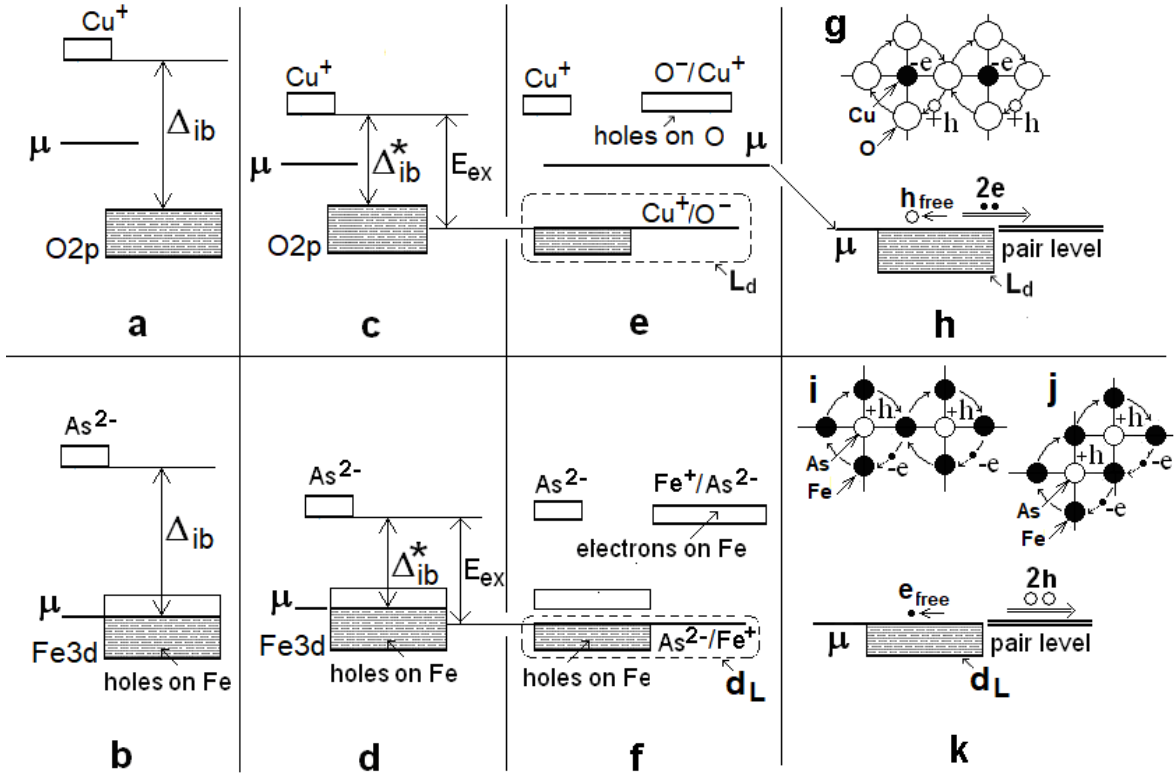

**Figure 3 a,b**, A simplified two-band scheme for the electronic structures of undoped cuprates (**a**) and ferropnictides (**b**). Energy $\Delta_{ib}$ required for interband transition is in both cases approximately the same, $\Delta_{ib} \sim 2$ eV. **c,d**, At a reduction of the interband gap to $\Delta_{ib} = \Delta^*_{ib} \leq E_{ex}$, two-particle transitions to and from become possible between two one-particle states (p electron + d hole) on the one hand, and an exciton two-particle state (d electron + p hole) on the other hand. **e**, Part of electrons from the oxygen $O^{2-}$ band pass into bound $Cu^+/O^-$ states (i.e., an electron on the central Cu cation and a hole distributed over four surrounding anions of O). Herewith, the bound oxygen holes occupy $O^-/Cu^+$ states. **f**, Part of holes from the Fe band pass into bound $As^{2-}/Fe^+$ states (i.e., a hole on the As anion and an electron distributed over four surrounding cations of Fe). Herewith, electrons occupy $Fe^+/As^{2-}$ states. Dashed lines in figures (**e,f**) delineate composite bands $L_d$ and $d_L$, in which electron states in cuprates (hole states in ferropnictides) are a superposition of band and exciton bound states. **g**, Two CT plaquettes centred on neighbouring Cu ions form an HL centre in cuprates. **h**, An "ionized" HL centre can exist, on which two electrons and one hole in cuprates form a bound state. As the result, free hole carriers will appear in an $L_d$ band. **i,j**, Two different types of CT plaquettes in ferropnictides. **k**, In ferropnictides, two holes and one electron can form a bound state. As the result, free electron carriers will appear in a $d_L$ band.

A similar transition into the state of a CT exciton insulator takes place at $\Delta^*_{ib} < E_{ex}$ in ferropnictides. Herewith, part of holes from the Fe3d band will pass into $As^{2-}/Fe^+$ exciton states, where $Fe^+$ denotes an additional electron on the Fe neighbouring the As ion (Fig. 3f). These additional electrons passing from the As to neighbouring Fe ions occupy $Fe^+/As^{2-}$ states.

Thus, in the CT phase both cuprates and ferropnictides can be considered as exciton insulators, where two-particle transitions to and from become possible between two one-particle states (p electron + d hole) on the one hand and an excitonic two-particle state (d electron + p hole) on the other hand. Now let us consider how such systems in the CT phase become conductors and from where free carriers emerge.

Let there be two hydrogen-like CT plaquettes centred on the nearest cations of Cu in cuprates (Fig. 3g) or on the nearest anions of As in ferropnictides (Fig. 3i,j). Such a pair of CT plaquettes represents a solid state analogue of the hydrogen molecule [7] and can be considered as an HL centre. In cuprates, only one type of HL centres is possible (Fig. 3g), whereas ferropnictides may feature two types corresponding to the arrangement of neighbouring As ions on one or on different (higher and lower) sides of the FeAs plane (Fig. 3i,j). Note that in the CT phase each pair of the nearest Cu ions together with enclosing O ions (in cuprates), as each pair of the nearest As ions together with enclosing Fe ions (in ferropnictides), plays the role of HL centres.

Similar to a molecule of $H_2$, on an HL centre two electrons and two holes can form a bound state due to the possibility of two singlet holes (in cuprates) or electrons (in ferropnictides), being in the space between the central ions, to be attracted to, correspondingly, two electrons or holes being on these ions. An additional decrease of energy, $\Delta E_{HL}$, in this case can be estimated from the ratio $\Delta E_{HL} \sim \Delta E_{H2}/\varepsilon_\infty^2 \approx 0.2$ eV, where $\Delta E_{H2} = 4.75$ eV is bond energy in the molecule of $H_2$, and $\varepsilon_\infty \approx 4.5$–5 (for cuprates) [6]. The chemical potential of bound pairs (the pair level in Fig. 3h,k) in cuprates and ferropnictides coincides with the top of the $L_d$ and $d_L$ bands, respectively, at $T = 0$.

Continuing the analogy with the $H_2$ molecule, we note that, apart from this molecule, a molecular ion $H_2^+$ can also exist. Also, similarly with the molecular ion, an "ionized" HL centre can exist, on which two electrons and one hole (in cuprates) or two holes and one electron (in ferropnictides) form bound states. That is to say, the HL centre can act as an acceptor (in cuprates) or donor (in ferropnictides). As the result, free hole carriers in the O2p band (in cuprates, Fig. 3h) and free electron carriers in the Fe3d band (in ferropnictides, Fig. 3k) will appear.

Thus, in the CT phase cuprates and ferropnictides are self-doped CT excitonic insulators, where, as the result of the interaction of band electrons or holes with HL centres, free carriers are generated. In cuprates, these are always holes; in ferropnictides, always electrons. These carriers are not associated with doping. Doping, as we will show below, is required only to form the CT phase in initially undoped cuprates and ferropnictides.

## 2 Features of phase diagrams of cuprate and ferropnictide HTSCs

To verify the proposed model, let us consider how doping deforms the electronic structures of undoped cuprates and ferropnictides to lead to the formation of regions of a CT phase in their basal planes, in which phase each Cu ion (in cuprates) or As ion (in ferropnictides) is the centre of a CT plaquette. This will enable us to determine the regions of dopant concentrations within which the percolation cluster of the CT phase (or, to be more exact, the percolation cluster of HL centres) can exist. We will show that these regions exactly coincide with the positions of the superconducting domes on the phase diagrams of all cuprates and ferropnictides, for which these diagrams are well established. Besides, we will show that this approach enables predicting with wonderful accuracy the positions of characteristic features on $T_c(x)$ curves inside the superconducting dome, as well as the sharp peaks on the curves for the dependences of the London depth of penetration $\lambda$ on $x$ for a number of compounds, which is indicative of a decrease in the density of pairs in the narrow ranges of dopant concentrations.

Let us recall here once again: an HL centre is a pair of CT plaquettes ($CuO_4$ or $FeAs_4$) centred on neighbouring ions of Cu or As. An aggregate of non-intersecting HL centres, in which all ions of O or Fe can be went around by passing from one ion of O or Fe to another nearest ion, we will call a cluster of HL centres in the basal plane of $CuO_2$ (or FeAs).

As the gap $\Delta_{ib}$ in undoped cuprates and ferropnictides is largely determined by Madelung energy $E_M$, we need to locally decrease the value of $E_M$ to form a CT plaquette centred on a given (Cu or As) ion. Hypothetically, this can be done by arranging additional localized charges of respective value and sign either on the central (Cu or As) ion or on one of four surrounding ions (of O or Fe). Interestingly, exactly this mechanism of decreasing $\Delta_{ib}$ appears to be realized in HTSCs under doping.

The insertion of a heterovalent dopant ion into the lattice is accompanied with the emergence of an additional carrier (electron or hole) in the basal plane or near it. The charge of this doped carrier, in accordance with the symmetry of the environment, is spread over the nearest ions, inducing on them a fractional charge $q^* \approx \pm|e|/4$ (Fig. 4). This figure presents various examples of electron and hole doping for particular compounds of cuprates and ferropnictides with different symmetries of the distribution of the doped carrier. These examples cover virtually all variants of doping.

At a hole doping a positive charge is induced on the nearest ions of O or As (Fig. 4a,b,f); at an electron doping, a negative charge, on four nearest ions of Cu or Fe (Fig. 4c–e). Note that in the case of YBa$_2$Cu$_3$O$_{6+x}$ (Fig. 4a), for charge $q^* \geq \pm|e|/4$ to emerge on apical oxygen ions, excess oxygen should occupy two successive positions in the chain.

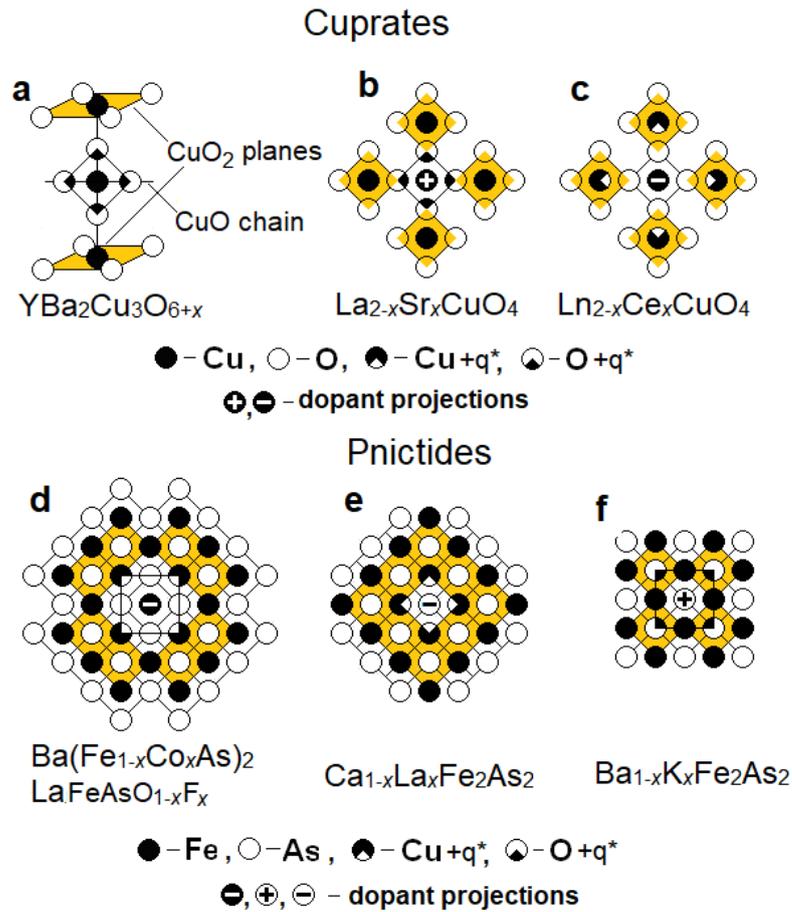

**Figure 4 Formation of CT plaquettes and trion complexes in cuprates and ferropnictides.** The CT plaquette is highlighted in yellow. The charge of this doped carrier, in accordance with the symmetry of the environment, is spread over the nearest ions, inducing on them a fractional charge $q^* \approx \pm|e|/4$ (white or black sectors). In case (**f**), charge $q^*$ is induced on the central ion of Fe. **a**, A dopant and a doped charge outside the basal plane (YBCO). **b**, Hole doping in LSCO. **c**, Electron doping in LCCO. **d**, Electron doping in BFCA and LFAOF ferropnictides. **e**, Electron doping in CLFA ferropnictides. **f**, Hole doping in BKFA ferropnictides.

In turn, each emerging induced charge $q^* \approx \pm|e|/4$ locally decreases the initial gap $\Delta_{ib}$ in its nearest vicinity down to a value $\Delta_{ib}^*$, which, according to estimates [7], enables an exciton transition with charge transfer between the Cu (As) ion in the basal plane and neighbouring O (Fe) ions. This corresponds to the formation of two (Fig. 4a) up to eight (Fig. 4d,e) CT plaquettes (highlighted in yellow) in the vicinity of the dopant projection.

As each CT plaquette formed by charge $q^*$ is bound to it by Coulomb attraction, the doped carrier self-localizes due to the formation of a trion complex, which includes the carrier $\pm|e|$ proper and its surrounding CT plaquettes. The localization of a doped carrier in the nearest vicinity of the dopant (at least at sufficiently low temperatures) has been confirmed by the results of works [8-12].

The localization boundary of the doped carrier is determined by the condition that the same charge $q^*\approx+|e|/4$ ($-|e|/4$) sufficient for CT plaquette formation be at this boundary per each anion (cation). In fact, we produce a trion complex where the doped carrier is bound to CT excitons that resonantly interact with the band states.

By condition, each pair of neighbouring CT plaquettes forms an HL centre (Fig. 3g,i,j). In turn, trion complexes get ordered in the basal plane, enabling the formation of percolation clusters of HL centres (or CT phase regions). The dopant concentration range corresponding to the existence of a percolation cluster of HL centres (the superconducting CT phase) can be readily determined for each particular compound. The localization regions of doped carriers (non-highlighted central plaquettes in Fig. 4) should be attributed to the overdoped (non-superconducting) phase.

Optimal doping will conform to the complete (without overlapping) filling of the plane by the patterns shown in Fig. 4. For cuprates (Fig. 4a–c) the optimal concentration should be achieved, respectively, at $x_{opt}$ =1, 1/5 and 1/5. For ferropnictides (Fig. 4d–f), respectively, $x_{opt}$ = 1/12, 2/9 and 2/5 (it is taken into account here that in cases of Fig. 4e,f an electron or a hole are doped into one of the two planes). These values for cuprates and ferropnictides agree well with the experimental values of the optimal concentrations for doping [13-19]. A slight excess of the presented values of the optimal concentrations with reference to the experimentally observed values is due to the impossibility to implement experimentally a dense packing of patterns in Fig. 4 without overlapping CT plaquettes. Herewith, the overlapping regions (delocalization of doped carriers) will have at an increased level of doping and correspond to the overdoped (metallic, non-superconducting) phase.

Note that in the 1111 system (LaFeAsO$_{1-x}$F$_x$), at the substitution of fluorine for oxygen, one electron is doped into the basal plane. The projection of the F ion onto the basal plane coincides with the position of the Co ion in Fig. 4d. As a consequence, the symmetry of doped charge distribution in LaFeAsO$_{1-x}$F$_x$ will be similar to that in Ba(Fe$_{1-x}$Co$_x$As)$_2$. Therefore, the regions of concentrations corresponding to a superconducting dome, as the values of the optimal concentrations in these materials, should coincide, which is consistent with the experiment [16,19].

Let us consider here the most demonstrable coincidences of the calculated ranges for the existence of HL centres' percolation clusters in various compounds with the positions of the superconducting domes on their phase diagrams.

Let us begin with the simplest case, $YBa_2Cu_3O_{6+x}$. Doping of the parent compound $YBa_2Cu_3O_6$ is performed by introducing excess oxygen $x$ into the plane of the chains. In the case when two consecutive positions in a chain are occupied with oxygen ions (Fig. 4a), an oxygen square is formed with one hole distributed over four oxygen ions of this square. Herewith, additional positive charges $q^*$ ($\approx +|e|/4$) emerge on the apical ions of oxygen, nearest to the plane ions of Cu, which results in the formation of CT plaquettes with the centre on these Cu ions. Thus, in $YBa_2Cu_3O_{6+x}$ a doped charge is localized outside the basal plane, and the trion complex formed under doping represents two CT plaquettes localized in the $CuO_2$ planes over different sides of the plane of the CuO chains (Fig. 4a).

To form a CT plaquette on a Cu ion in the basal plane, it is required to fill two consecutive oxygen positions in the chain over/under it (Fig. 4a). The concentration of such oxygen pairs (and, therefore, CT plaquettes) is $x^2$, and the percolation cluster of HL centres in the $CuO_2$ plane (Fig. 3g) will exist at an excess of the site percolation threshold in the square $CuO_2$ lattice, which is $x_p = 0.593$ [20]. Thus, the percolation cluster of a CT phase will exist at $x^2 > 0.593$, i.e., at $x > 0.77$. Herewith, the region of optimal doping for $YBa_2Cu_3O_{6+\delta}$ is within the interval of $0.77 < \delta < 1$, in accordance with the experiment (see, e.g., Fig. 2a in [13]). In the region of $x < 0.77$ the existence of CT-phase finite clusters of various sizes is possible [21]. In this case, superconductivity emerges owing to the Josephson coupling between clusters.

Note that YBCO represents the only known example of HTSCs, where CT phase can cover the entire basal plane under doping. In the case of, e.g., another two-plane cuprate $Ba_2Sr_2CaCu_2O_{8+x}$, where $0 < x < 2$, an inhomogeneous filling of excess oxygen positions leads to various charges $q^*$ on the apical ions of oxygen, a consequence of which is the co-existence of underdoped and overdoped regions, together with optimally doped ones, in one basal plane. Cases of three- or four-plane cuprates require special consideration owing to the occurrence of infinite (superconducting) layers of $CaCuO_2$ with an unknown character of doping in their structure.

In the cases when doped charges are in the basal plane they form four up to eight CT plaquettes depending on the symmetry of the environment (Fig. 4b–f). Formation of a percolation network of HL centres that represent pairs of neighbouring CT plaquettes (Fig. 3g,i,j) is possible at an ordered arrangement of trion complexes (or projections of dopants) into the superlattice with parameter $l_D$ with site occupation $0.593 \leq \nu \leq 1$ (here 0.593 is the site percolation threshold in the square lattice).

Let us have a domain in the basal plane where the trion centres are randomly distributed over the sites of a superlattice with parameter $l_D$. The dopant concentration $x_{max} = 1/l_D^2$ corresponding to the full filling of superlattice sites will be taken for the upper boundary of the optimal doping region. This boundary corresponding to the maximal concentration of HL centres also corresponds, according to our assumption, to the maximum $T_c$. For the lower boundary, $x_{min} = 0.593/l_D^2 \approx 0.6/l_D^2$, we will conventionally take a dopant concentration corresponding to the site percolation threshold in a square superlattice with parameter $l_D$ [20]. This choice is determined by the fact that the existence of physically significant domains with the percolation network of dopant projections spaced by a distance $l_D$ from each other is possible only at $0.593/l_D^2 < x \leq 1/l_D^2$.

This does not mean that the percolation cluster on a lattice with parameter $l_D$ should occupy the entire crystal, but only that such domains will exist within only this concentration range. Herewith, a superconductivity in the entire crystal can emerge owing to the Josephson coupling between such domains. A particular shape of the curve of $T_c(x)$ is determined by the competition of order and disorder, in other words, by the possibility of realizing various types of trion ordering in this crystal structure. As examples, let us consider cuprate La$_{2-x}$Sr$_x$CuO$_4$ and ferropnictide Ca$_{1-x}$La$_x$Fe$_2$As$_2$.

In the case of La$_{2-x}$Sr$_x$CuO$_4$, the doped hole emerging at the substitution of Sr$^{2+}$ for La$^{3+}$ is in the CuO$_2$ plane (Fig. 5) and is distributed over four oxygen ions pertaining to the oxygen octahedron adjacent to Sr ion [8,9]. Each of four fractional charges $q^*$ on oxygen ions forms CT plaquette in the next ion square centred on the nearest Cu cation (highlighted in yellow in Fig. 5). It is readily seen that only two variants of the relative arrangement of two nearest projections of Sr on the CuO$_2$ plane are possible so that they could form an HL centre (Fig. 5a,c). These cases correspond to two possible distances between them, $l_D = 3a$ and $l_D = a\sqrt{5}$ ($a$, lattice parameter). The pairs of neighbouring CT plaquettes that form HL centres are highlighted in orange.

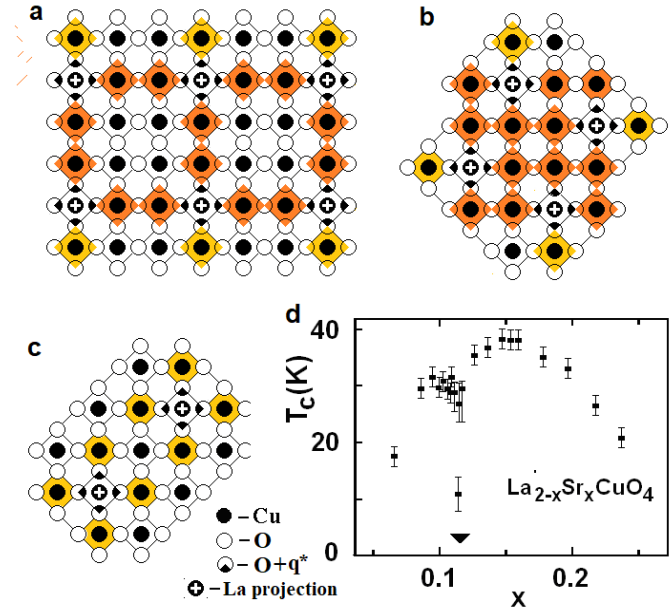

**Figure 5 Formation of HL centres from pairs of CT plaquettes in $La_{2-x}Sr_xCuO_4$ at various distances between trion centres: a**, $l_D = 3a$; **b**, $l_D = a\sqrt{8}$; **c**, $l_D = a\sqrt{5}$ . Circles with black sectors are oxygen ions carrying an additional charge $q^*$. Pairs of CT plaquettes forming HL centres are highlighted in orange. CT plaquettes not forming HL centres are highlighted in yellow; **d**, The phase diagram of $La_{2-x}Sr_xCuO_4$ [14].

Note that in an intermediate case when the distance between the Sr projections is $l_D = a\sqrt{8}$ (Fig. 5b), no HL centres are formed. In accordance with two different variants of dopant projections' arrangement, the phase diagram of $La_{2-x}Sr_xCuO_4$ is expected to have two regions of optimal doping at $0.066 < x < 0.11$ and at $0.12 < x < 0.2$ (corresponding to the ordered arrangement of Sr projections onto 3x3 and $\sqrt{5}$x$\sqrt{5}$ lattices). Note that the experimental value of the upper optimal concentration, $x = 0.15$ (optimal in the sense of the magnitude of $T_c$) differs from the expected value, $x = 1/5$, though a jumplike decrease of the volume of the superconducting phase is observed namely at $x = 1/5$ [22]. We explain this discrepancy by the formation of clusters of normal metal, which begins at $x > 0.15$ [22]. However, if we succeed in providing for Sr ordering at concentrations exceeding $x = 0.15$, the range of maximal $T_c$ can be extended up to $x = 0.2$ [23]. A similar slight discrepancy between the expected and experimental values of $x_{opt}$ takes place for other cuprate and ferropnictide compounds, too; this is also associated with incomplete ordering and the beginning formation of overdoped phase domains. Within the interval of $0.11 < x < 0.12$, clusters of CT phase cannot exist. This is the so-called 1/8 anomaly, which, however, takes place not at $x = 0.125$ but, according to this consideration, at $x = 0.115$, in total agreement with experimental data [14,24].

In electron-doped ferropnictide $Ca_{1-x}La_xFe_2As_2$ the substitution of La for Ca leads to the emergence of an additional electron in one of the two FeAs planes that imparts charge $q^*$ ($\approx -|e|/4$) to each of four Fe ions nearest to the projection of La (Fig. 6a). As the result, eight CT plaquettes form around these Fe ions.

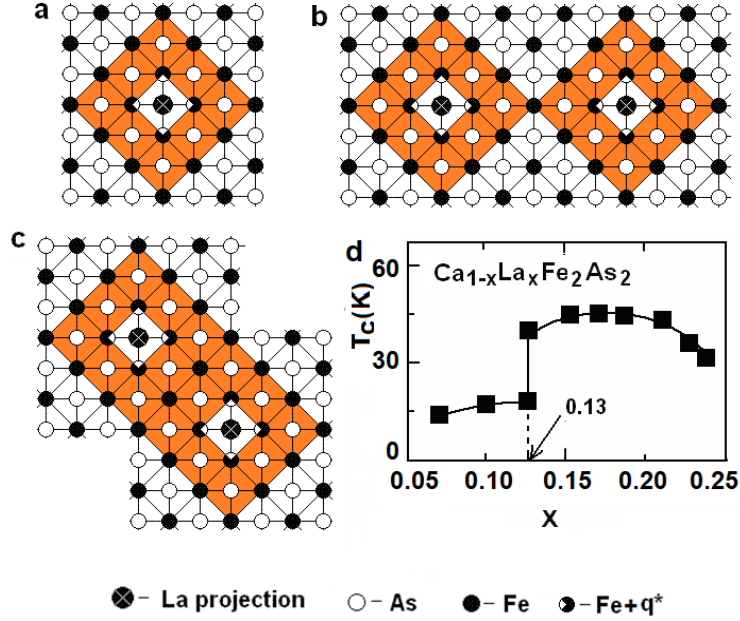

**Figure 6 a**, Formation of CT plaquettes in $Ca_{1-x}La_xFe_2As_2$. Eight CT plaquettes (yellow highlights) form around a La projection in one of the FeAs planes; open circles, As; filled circles, Fe; filled minus circles, Fe ions with charge $q^* = -e/4$. **b**, Formation of a percolation cluster of CT plaquettes on the $\sqrt{18}\times\sqrt{18}$ superlattice. **c**, Formation of a percolation cluster of CT plaquettes on the 3×3 superlattice; pairs of CT plaquettes forming HL centres are highlighted in orange. **d**, The phase diagram of $Ca_{1-x}La_xFe_2As_2$ [17].

The maximal possible distance between dopant projections to form a percolation cluster of HL centres is $l_D = \sqrt{18}$ (Fig. 6b). The percolation threshold at the ordering of dopant projections to a $\sqrt{18}\times\sqrt{18}$ superlattice corresponds to a concentration of $x \approx 2\cdot 0.033=0.066$ (with account for the fact that only each second La ion dopes an electron to the FeAs plane). Herewith, the bonding between the trion complexes will be performed via HL centres of the first type (Fig. 3i).

Optimal doping as seen in Fig. 6c is in correspondence with the complete ordering of dopant projections to a 3×3 superlattice. A concentration corresponding to an optimal doping is $x = 0.22$; respectively, a concentration that conforms to the percolation threshold at this lattice is $x = 0.132$. In this case, the percolation cluster will include HL centres of the second type only (Fig. 4j).

All three concentration values (0.066, 0.13 and 0.22) coincide with the boundaries for the low- and high-temperature superconductivity regions on the experimental phase diagram of $Ca_{1-x}La_xFe_2As_2$ (Fig. 2g [17]).

Using Fig. 4, which shows variants of trion complexes for various types of doping, we can also find the positions of the superconducting domes on the phase diagrams of other cuprates and ferropnictides.

Recent years have witnessed publications the authors of which observed in ferropnictides $BaFe_2(As_{1-x}P_x)_2$ [25,26] and $Ba(Fe_{1-x}Co_xAs)_2$[27] the sharp peaks on the curves for the dependences of the London depth of penetration $\lambda$ on $x$, which is indicative of a decrease in the density of pairs in the narrow range of dopant concentrations (Fig. 7a). According to the proposed model, such a decrease should be related to the decrease in the power of the percolation cluster of HL centres in a certain range of dopant concentrations. Let us consider this in greater detail.

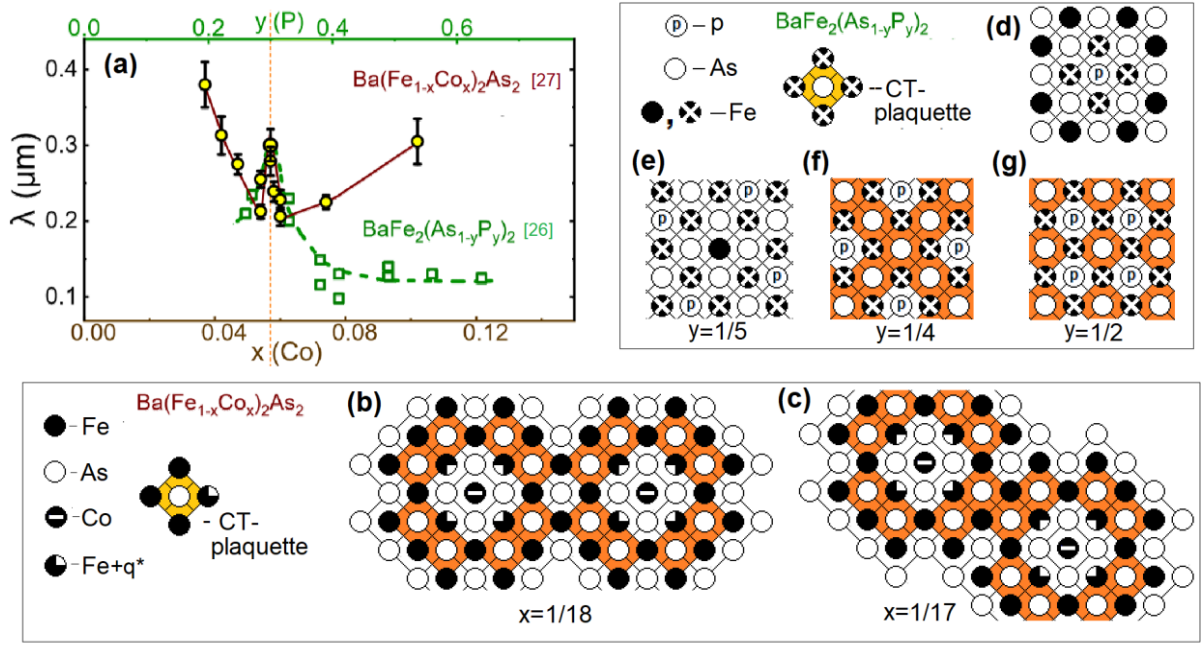

**Figure 7 a,** Dependences of the London penetration depth $\lambda$ in $Ba(Fe_{1-x}Co_xAs)_2$ and $BaFe_2(As_{1-y}P_y)_2$ on doping [from 26,27]; **b,c,** Formation of clusters of HL centres in $Ba(FeAs_{1-y}P_y)_2$ at different dopant concentrations; **b,** The highest concentration of Co at which the number of HL centres (highlighted in orange) in type I clusters can be maximized due to ordering, $x_{Co1} = 1/18$; **c,** The lowest concentration of Co at which the number of HL centres (highlighted in orange) in type II clusters can be maximized due to ordering, $x_{Co2} = 1/17$; **d-g,** Formation of clusters of HL centres in $Ba(Fe_{1-x}Co_xAs)$ at different dopant concentrations. **d,** Decrease in the gap $\Delta_{ib}$ for transitions of holes from the Fe ions nearest to P (highlighted by crosses) to the neighbouring ions of As to a value of $\Delta_{ib}^*$; **e,** At $x = 1/5$, no CT plaquettes are formed (the number of Fe ions with crosses around the As ion is less than 4); **f,** At $x = 1/4$, in the ordered (2×2) phase there is a percolation cluster of HL centres (highlighted in orange) in which CT plaquettes touch one another by their sides; **g,** At $x = 1/2$, in the ordered $(\sqrt{2} \times \sqrt{2})$ phase there is a percolation cluster of HL centres (highlighted in orange) in which CT plaquettes touch one another by their angles.

In the compound Ba(Fe$_{1-x}$Co$_x$As)$_2$, clusters of HL centres form at the touch of trion complexes (by their angles or sides, Fig. 7). At low concentrations of Co, clusters of HL centres form mainly due to the touch of trion complexes by their angles (type I clusters), whereas at high concentrations, due to the touch by their sides (type II clusters). At $x = 0$ and $x > 0.11$, HL centres are absent. In the former case, due to the absence of trion complexes; in the latter, to the overlap of the doped carriers' localization regions, as the result of which these complexes cease to exist. At an increase of the concentration of Co starting from $x = 0$, as well at a decrease from $x < 0.11$, the number of HL centres will increase. The highest concentration of Co, at which the number of HL centres in a cluster could be maximized due to the ordering of trion complexes touching one another by their angles, is $x_{Co1} = 1/18$ (Fig. 7b). The lowest concentration of Co, at which the number of HL centres in a cluster could be maximized due to the ordering of trion complexes touching one another by their sides, is $x_{Co2} = 1/17$ (Fig. 7c). Correspondingly, the maxima of the concentration of the superconducting pairs $\rho_S(0)$, which is determined by the concentration of HL centres, and the minima of the London penetration depth $\lambda_L(0)$ should be observed at the same values (1/18 and 1/17) [27].

In between $x_{Co1}$ and $x_{Co2}$, any cluster will include a smaller number of HL centres, because it is impossible in this range to order all trion complexes forming them. Thus, the concentration of the superconducting pairs will have a minimum at $(x_{Co1} + x_{Co2})/2 \approx 0.057$, and the London penetration depth will have a maximum at the same value of x, which corresponds to the experiment [27]. Note that if the arrangement of dopants could be ordered at larger concentrations, e.g., at $x = 1/13$, a similar minimum of $\lambda_L(0)$ could be observed at that value of $x$, too.

In the compound BaFe$_2$(As$_{1-x}$P$_x$)$_2$, superconductivity emerges as the result of the isovalent substitution of P ions for As (Fig. 7a). Herewith, as we believe, ions of P locally deform the electronic structure of the parent compound BaFe$_2$As$_2$ such that to decrease the gap $\Delta_{ib}$ for transitions of holes from the Fe ions nearest to P to the neighbouring ions of As to a value of $\Delta_{ib}^*$ (in Fig. 7b, these Fe ions are highlighted in grey with crosses).

To form a CT plaquette centred on an As ion, it is required that ions of P be at certain distances: $2a$ or $a\sqrt{2}$. Note that at $x = 0.2$ (at an ordered arrangement, Fig. 7) there will still be no CT plaquettes, and, therefore, no HL centres responsible for superconductivity (Fig. 7c). At a further concentration increase, CT plaquettes and, along with them, HL centres (mainly of the second type, Fig. 3j) do appear, as the result of which superconductivity emerges. At $x = 0.25$ (at an ordered arrangement) the number of CT plaquettes reaches a maximum (Fig. 7d). This concentration will correspond to the pair density maximum and, correspondingly, to a local minimum of λ. As the value of $x > 0.25$ is exceeded, the percolation cluster of HL centres partially falls apart and reformats into another cluster with HL centres of the first type (Fig. 3i), the maximal number of which corresponds to $x = 0.5$ (the ordering into a lattice $\sqrt{2} \times \sqrt{2}$, Fig. 7e). The site percolation threshold in this lattice is $x_p = 0.593 \times 0.5 \approx 0.3$. Therefore, starting from $x = 0.3$, the number of HL centres and the density of pairs begin to rise, and λ drops down to $x = 0.5$, which totally agrees with the experiment.

### 3 Mechanism of superconducting pairing

Here we propose a mechanism of superconducting pairing genetically inherent in such a system. The basis of the proposed mechanism is the interaction of electrons with HL centres. As the works [28-33] have shown, the account for the scattering processes with intermediate virtual bound states that are in vicinity of $E_F$ can lead to a strong renormalization of the efficient interelectron interaction, capable of providing for high $T_c$ in the system.

Let us consider how this mechanism could work in cuprates and ferropnictides.

According to the above said, incoherent transport in the CT phase of cuprates (ferropnictides) at $T > 0$ is performed by hole (electron) carriers emerging at the occupation of HL centres by electrons (holes). The concentration of free carriers $n$ will be determined from the condition of equality of the chemical potentials for pairs on the HL centres and of electrons in O2p (or holes in Fe3d) band.

For small concentrations when the occupation of separate HL centres can be considered to be independent one of another, the change of the chemical potential for pairs due to transition of part of electrons (holes) from a band to HL centres can be presented as [34]:

$$\Delta\mu_p = -\frac{T}{2} \ln[f(n)], \quad (1)$$

where $n$ is the two-dimensional concentration of formed free carriers, and $f(n)$ is some function of occupation of HL centres. Herewith, the change of the chemical potential for band carriers $\Delta\mu \approx -n/N(0)$, where $N(0)$ is the density of states on the Fermi level in a band. Equating the chemical potential changes of pairs and band electrons, we find in an explicit form the dependence of hole concentrations $n$ on temperature:

$$T \approx 2n/N(0)\ln[f(n)]. \quad (2)$$

Thus, in the CT phase at low $T$, $n \propto T$. At high $T$, the concentration tends to a saturation, when the population of copper sites tends to unity (i.e., all HL centres are occupied).

As it follows from (2), at $T = 0$ the pair level is empty, and incoherent transport is impossible owing to the absence of free carriers. At the same time, such a system where an electron and a hole are present in each CT plaquette enables coherent transport, when all carriers of the same sign move coherently, as a single whole, e.g., a superconducting condensate. The latter is possible given a superconducting pairing. We believe that in the proposed pattern based on the formation of local CT excitons and HL centres such an interelectron attraction in the CT phase emerges due to the formation of a bound state of two electrons (holes), getting on the central Cu (As) ions of an unoccupied HL centre, and two holes (electrons), with necessity emerging herewith on the surrounding O (Fe) ions. The value of the pairing interaction will be determined by the coupling energy of pairs of electrons and holes on HL centres.

Note that under certain conditions we can observe in such systems a change of sign of carriers in the Hall effect at a superconducting transition. This change of carrier sign in some cuprates and ferropnictides is indeed observed in the experiment [35,36]. In addition, it should also be noted that, as each state in $L_d$ and $d_L$ bands (Fig. 3e,f) is a superposition of band and exciton states, all electrons will be involved in pairing.

**Conclusion**

Thus, we propose a doping model common for cuprates and ferropnictids, which assumes the formation, around a doped carrier, of a trion complex consisting of the doped carrier and enclosing CT excitons. Each trion complex occupies an area of several neighbouring cells in the basal plane that form a geometric figure whose shape depends on the lattice symmetry, as well as the type and position of dopant. For each material from the above families, this shape is unambiguously determined by a simple common rule.

In a certain range of dopant concentrations, trion complexes interacting one with another form percolation clusters of a CT phase. With the geometry of trion complexes known, these ranges can be readily determined for each compound. Calculations showed the concentration ranges corresponding to the superconducting dome on the phase diagrams of cuprates and ferropnictides to coincide with those for the formation of a CT-phase percolation cluster. This approach also enabled us to describe fine features of the phase diagrams of cuprates and pnictides, including the 1/8 anomaly in $La_{2-x}Sr_xCuO_4$, sharp rises of $T_c(x)$ в $YBa_2Cu_3O_{6+x}$ and $Ca_{1-x}La_xFe_2As_2$, as well as narrow peaks of $\lambda_L(x)$ in $Ba(Fe_{1-x}Co_xAs)_2$ and $BaFe_2(As_{1-y}P_y)_2$.

These results can serve as confirmation of the proposed model and at the same time indicate the incorrectness of the standard approach, which considers cuprates and ferropnictides as spatially homogeneous systems with carrier concentrations determined by the doping level. The fact that the proposed method of constructing phase diagrams proved similarly successful both for cuprates and ferropnictides serves as a serious argument in favour of the common nature of their normal state and gives grounds to attribute these materials in the HTSC phase to CT exciton insulators. In the case of cuprates, the use of the proposed approach to the analysis of the transformation of the electronic structure with doping allowed us to explain a number of their anomalies: Fermi arcs, "large" and "small" pseudogaps, etc. [1].

Accepting that high temperature superconductivity emerges in the CT phase, it can be assumed that the very mechanism of superconducting pairing is determined by special properties of this phase, in which electron states are a superposition of band and exciton states. This enables considering the basal planes of doped cuprates and ferropnictides in the CT phase as one more type of structures (in addition to one-dimensional Little chains [37] and Ginzburg sandwiches [38]), where the exciton mechanism of superconductivity can be realized.

## Acknowledgements

This work was supported by Program of the Presidium of the Russian Academy of Sciences "Fundamental problems of high-$T_c$ superconductivity".

## Author contributions

Both authors contributed to the work presented in this paper. K.M. conceived the original idea and designed the model. O.I. provided critical feedback, analyzed previously conducted experiments from literature and helped shape the research. Both K.M. and O.I. discussed the results and contributed to the final manuscript.

## Competing interests

The authors declare no competing interests.

# References

1. K.V. Mitsen and O.M. Ivanenko, *Towards the issue of the nature of Fermi surface, pseudogaps and Fermi arcs in cuprate HTSCs*, J. Alloys Compd **791**, 30 (2019).

2. D. Allender, J. Bray, and J. Bardeen, *Model for an exciton mechanism of superconductivity.* Phys. Rev. B **7**, 1020 (1973).

3. D.J. Singh and M.-H. Du, *Density functional study of $LaFeAsO_{1-x}F_x$: A low carrier density superconductor near Itinerant magnetism*, Phys. Rev. Lett. **100**, 237003 (2008).

4. M. Nohara and K. Kudo, *Arsenic chemistry of iron-based superconductors and strategy for novel superconducting materials,* Adv. Phys. X **2(2)**, 450-461 (2017).

5. C.M. Varma, S. Schmitt-Rink, and E. Abrahams, *Charge transfer excitations and superconductivity in "ionic" metals,* Sol. St. Commun. **88**, 847 (1993).

6. H. Romberg, M. Alexander, N. Nucker, P. Adelmann, and J. Fink, *Electronic structure of the system $La_{2-x}Sr_xCuO_{4+d}$*, Phys. Rev. B **42,** 8769 (1990).

7. K.V. Mitsen and O.M. Ivanenko, *Phase diagram of $La_{2-x}M_xCuO_4$ as the key to understanding the nature of high-$T_c$ superconductors*, Phys. Usp. **47,** 493 (2004).

8. D. Haskel, V. Polinger, and E.A. Stern, *Where do the doped holes go in $La_{2-x}Sr_xCuO_4$? A close look by XAFS*, AIP Conf. Proc. **483**, 241 (1999).

9. P.C. Hammel, B.W. Statt, R.L. Martin, F.C. Chou, D.C. Johnston, and S.-W. Cheong, *Localized holes in superconducting lanthanum cuprate,* Phys. Rev. B **57**, 712 (1998).

10. H. Wadati, I. Elfimov, and G.A. Sawatzky, *Where are the extra d electrons in transition-metal-substituted iron pnictides?* Phys. Rev. Lett. **105**, 157004 (2010).

11. T. Berlijn, C.H. Lin, W. Garber, and W. Ku, *Do transition-metal substitutions dope carriers in iron-based superconductors?,* Phys. Rev. Lett. **108**, 207003 (2012).

12. G. Levy, R. Sutarto, D. Chevrier, T. Regier, R. Blyth, J. Geck, S. Wurmehl, L. Harnagea, H. Wadati, T. Mizokawa, I. S. Elfimov, A. Damascelli, and G.A. Sawatzky, *Probing the role of Co substitution in the electronic structure of iron pnictides*, Phys. Rev. Lett. **109,** 077001 (2012).

13. K. Segawa and Y. Ando, *Transport anomalies and the role of pseudogap in the 60-K phase of $YBa_2Cu_3O_{7-\delta}$,* Phys. Rev. Lett. **86**, 4907 (2001).

14. K.I. Kumagai, K. Kawano, I. Watanabe, K. Nishiyama, and K. Nagamine, *Magnetic order and evolution of the electronic state around x= 0.12 in $La_{2-x}Ba_xCuO_4$ and $La_{2-x}Sr_xCuO_4$,* J. Supercond. **7**, 63 (1994).

15. Y. Krockenberger, J. Kurian, A. Winkler, A. Tsukada, M. Naito, and L. Alff, *Superconductivity phase diagrams for the electron-doped cuprates $R_{2-x}Ce_xCuO_4$ (R= La, Pr, Nd, Sm, and Eu)*, Phys. Rev. B **77**, 060505 (2008).

16. Lei Fang, Huiqian Luo, Peng Cheng, Zhaosheng Wang, Ying Jia, Gang Mu, Bing Shen, I.I. Mazin, Lei Shan, Cong Ren, and Hai-Hu Wen, *Roles of multiband effects and electron-hole asymmetry in the superconductivity and normal-state properties of $Ba(Fe_{1-x}Co_x)_2As_2$*, Phys. Rev. B **80**, 140508 (2009).

17. Y. Sun, W. Zhou, L.J. Cui, J.C. Zhuang, Y. Ding, F.F. Yuan, J. Bai, and Z.X. Shi, *Evidence of two superconducting phases in $Ca_{1-x}La_xFe_2As_2$*, AIP Adv. **3**, 102120 (2013).

18. M. Rotter, M. Pangerl, M. Tegel, and D. Johrendt, *Superconductivity and crystal structures of $(Ba_{1-x}K_x)Fe_2As_2$ (x=0–1)*, Angew. Chem. Int. Ed. **47**, 7949 (2008).

19. Y. Kamihara, T. Watanabe, M. Hirano, and H. Hosono, *Iron-based layered superconductor $La[O_{1-x}F_x]FeAs$ (x=0.05−0.12) with Tc = 26 K*, J. Am. Chem. Soc. **130**, 3296 (2008).

20. J.L. Jacobsen, *High-precision percolation thresholds and Potts-model critical manifolds from graph polynomials*, J. Phys. A: Math. Theor. **47**, 135001 (2014).

21. S.Y. Gavrilkin, O.M. Ivanenko, V.P. Martovitskiĭ, K.V. Mitsen, and A.Y. Tsvetkov, *Percolation nature of the 60-K to 90-K phase transition in $YBa_2Cu_3O_{6+\delta}$*, J. Exp. Theor. Phys. **110**, 783 (2010).

22. H. Takagi, R.J. Cava, M. Marezio, B. Batlogg, J.J. Krajewski, W.F. Peck, Jr., P. Bordet, and D.E. Cox, *Disappearance of superconductivity in overdoped $La_{2-x}Sr_xCuO_4$ at a structural phase boundary,* Phys. Rev. Lett. **68**, 3777 (1992).

23. K. Ikeuchi, K. Isawa, K. Yamada, T. Fukuda, J. Mizuki, S. Tsutsui and A.Q.R. Baron, *Growth, characterization and application of single-crystal $La_{2-x}Sr_xCuO_4$ having a gradient in Sr concentration*, Jap. J. Appl. Phys., **45**, 1594–1601 (2006).

24. T. Nagano, Y. Tomioka, Y. Nakayama, K. Kishio, and K. Kitazawa, *Bulk superconductivity in both tetragonal and orthorhombic solid solutions of $(La_{1-x}Sr_x)_2CuO_{4-\delta}$*, Phys. Rev. B **48**, 9689 (1993).

25. K. Hashimoto, *et al.*, *A sharp peak of the zero-temperature penetration depth at optimal composition in $BaFe_2(As_{1-x}P_x)_2$*. Science **336**, 1554 (2012).

26. Y. Lamhot, A. Yagil, N. Shapira, S. Kasahara, T. Watashige, T. Shibauchi, Y. Matsuda, and O. M. Auslaender, *Local characterization of superconductivity in $BaFe_2(As_{1-x}P_x)_2$*, Phys. Rev. B **91**, 060504(R) (2015).

27. K. Joshi, N. Nusran, M. Tanatar, K. Cho, S. Bud'ko, P. Canifeld, R. Fernandes, A. Levchenko, and R. Prozorov, *Quantum phase transition inside the superconducting dome of $Ba(Fe_{1-x}Co_x)_2As_2$ from diamond-based optical magnetometry*, arXiv:1903.00053

28. G.M. Eliashberg, *Possible mechanism of the superconductivity and of a resistance varying linearly with T*, JETP Lett. **46**, S81-S84 (1987).

29. B.A. Volkov and V.V. Tugushev, *Electron mechanism for the formation of heavy charged and neutral bosons in systems with variable valence*, JETP Lett. **46**, 245 (1987)].

30. P.I. Arseyev, *A possible mechanism for high-T superconductivity*, JETP **74**, 667 (1992).

31. E. Simanek, *Superconductivity at disordered interfaces*, Solid State Commun. **32**, 731 (1979)

32. C.S. Ting, D.N. Talwar, and K.L. Ngai, *Possible mechanism of superconductivity in metal–semiconductor eutectic alloys*, Phys. Rev. Lett. **45**, 1213 (1980).

33. H.-B. Schuttler, M. Jarrell, and D.J. Scalapino, *Superconducting Tc enhancement due to excitonic negative-U centers: A Monte Carlo study*. Phys. Rev. Lett. **58**, 1147 (1987).

34. I.O. Kulik, *Electronic transfer of local pairs and superconductivity in metal oxide compounds*, Int. J. Modern Phys. B **2(05)**, 851–865 (1988).

35. R. Jin and H.R. Ott, *Hall effect of $YBa_2Cu_3O_{7-d}$ single crystals*, Phys. Rev. B **57,** 13872 (1998).

36. L.M. Wang, Un-Cheong Sou, H.C. Yang, L.J. Chang, Cheng-Maw Cheng, Ku-Ding Tsuei, Y. Su, Th. Wolf, and P. Adelmann, *Mixed-state Hall effect and flux pinning in $Ba(Fe_{1-x}Co_x)_2As_2$ single crystals (x = 0.08 and 0.10)*, Phys. Rev. B **83**, 134506 (2011).

37. W.A. Little, *Possibility of Synthesizing an Organic Superconductor*, Phys. Rev. A **134**, 6 (1964).

38. V.L. Ginzburg, *Concerning surface superconductivity*, J. Exp. Theor. Phys. **47**, 1549 (1964).

39.